\documentclass[twocolumn]{aastex631} 

\usepackage{amsmath}	
\usepackage{amssymb}	
\usepackage{wasysym}
\usepackage{multirow}
\usepackage[flushleft]{threeparttable}
\usepackage{booktabs}
\usepackage{booktabs}
\usepackage{tablefootnote}

\newcolumntype{Y}{>{\centering\arraybackslash}X}
\usepackage{array}  
\newcommand{\PSUAA}{Department of Astronomy \& Astrophysics, 525 Davey Laboratory, The Pennsylvania State University, University Park, PA, 16802, USA}
\newcommand{\PSUCEHW}{Center for Exoplanets \& Habitable Worlds, 525 Davey Laboratory, The Pennsylvania State University, University Park, PA 16802, USA}

\newcommand{\Caltech}{Department of Astronomy, California Institute of Technology, Pasadena, CA 91125, USA}
\newcommand{\JHU}{Department of Physics \& Astronomy, Bloomberg Center, Johns Hopkins University, Baltimore, MD 21218, USA}
\newcommand{\Macquarie}{School of Mathematical and Physical Sciences, Macquarie University, Balaclava Road, North Ryde, NSW 2109, Australia}

\newcommand{\JPL}{Jet Propulsion Laboratory, California Institute of Technology, 4800 Oak Grove Drive, Pasadena, CA 91109, USA}
\newcommand{\MITEAPS}{Department of Earth, Atmospheric, and Planetary Sciences, Massachusetts Institute of Technology, Cambridge, MA 02139, USA}
\newcommand{\MITKavli}{Kavli Institute for Astrophysics and Space Research, Massachusetts Institute of Technology, Cambridge, MA 02139, USA}

\newcommand{\Carnegie}{Earth and Planets Laboratory, Carnegie Institution for Science, 5241 Broad Branch Road, NW, Washington, DC 20015, USA}

\newcommand{\Princeton}{Department of Astrophysical Sciences, Princeton University, 4 Ivy Lane, Princeton, NJ 08540, USA}

\newcommand{\Tsinghua}{Department of Astronomy, Tsinghua University, Beijing 100084, China}
\newcommand{\FlatironCCA}{Center for Computational Astrophysics, Flatiron Institute, 162 Fifth Avenue, New York, NY 10010, USA}

\newcommand{\UCO}{UC Observatories, University of California, Santa Cruz, CA 95064, USA}

\newcommand{\WMKO}{W.\ M.\ Keck Observatory, 65-1120 Mamalahoa Hwy, Kamuela, HI 96743, USA}
\newcommand{\SSL}{Space Sciences Laboratory, University of California, Berkeley, CA 94720, USA}
\newcommand{\UH}{Institute for Astronomy, University of Hawai‘i, 2680 Woodlawn Drive, Honolulu, HI 96822, USA}
\newcommand{\UCB}{Department of Astronomy, 501 Campbell Hall, University of California, Berkeley, CA 94720, USA}
\newcommand{\UCLA}{Department of Physics \& Astronomy, University of California Los Angeles, Los Angeles, CA 90095, USA}
\newcommand{\nexsci}{NASA Exoplanet Science Institute/Caltech-IPAC, California Institute of Technology, Pasadena, CA
91125, USA}
\newcommand{\COO}{Caltech Optical Observatories, California Institute of Technology, Pasadena, CA 91125, USA}
\newcommand{\Sydney}{Sydney Institute for Astronomy (SIfA), School of Physics, University of Sydney, NSW 2006, Australia}
\newcommand{\Kansas}{Department of Physics and Astronomy, University of Kansas, Lawrence, KS, USA}
\newcommand{\Warwick}{Physics Department, University of Warwick, Coventry CV4 7AL, United Kingdom}
\newcommand{\Yale}{Department of Astronomy, Yale University, New Haven, CT 06520, USA}

\newcommand{\Schmidt}{Astrophysics \& Space Institute, Schmidt Sciences, New York, NY 10011, USA}
\newcommand{\Amsterdam}{Anton Pannekoek Institute for Astronomy, University of Amsterdam, Science Park 904, 1098 XH Amsterdam, The Netherlands}
\newcommand{\ND}{Department of Physics and Astronomy, University of Notre Dame, Notre Dame, IN 46556, USA}
\newcommand{\Geneva}{Observatoire Astronomique de l'Université de Genève, Chemin Pegasi 51, 1290 Versoix, Switzerland}
\newcommand{\WDRC}{Center for Solar-Stellar Connections, White Dwarf Research Corporation,  9020 Brumm Trail, Golden, CO 80403, USA}
\newcommand{\CEA}{Universit\'e Paris-Saclay, Universit\'e Paris Cit\'e, CEA, CNRS, AIM, F-91191, Gif-sur-Yvette, France}
\newcommand{\SAC}{Stellar Astrophysics Centre, Aarhus University, Ny Munkegade 120, DK-8000 Aarhus C, Denmark}
\newcommand{\PortoIAC}{Instituto de Astrof\'{i}sica e Ci\^{e}ncias do Espaço, Universidade do Porto, Rua das Estrelas, 4150-762 Porto, Portugal}
\newcommand{\Porto}{Departamento de F\'{i}sica e Astronomia, Faculdade de Ci\^{e}ncias da Universidade do Porto, Rua do Campo Alegre, s/n, 4169-007 Porto,
Portugal}
\newcommand{\UNSW}{School of Physics, University of New South Wales, NSW 2052, Australia}
\newcommand{\ASTROTHREED}{ARC Centre of Excellence for All Sky Astrophysics in Three Dimensions (ASTRO-3D)}

\turnoffediting

\shorttitle{Asteroseismology of $\sigma$ Dra}
\shortauthors{Hon et al.}

\graphicspath{{./}{figures/}}

\begin{document}

\title{Asteroseismology of the Nearby K-Dwarf $\sigma$ Draconis using the Keck Planet Finder and TESS} %

\author[0000-0003-2400-6960]{Marc Hon}
\affiliation{\MITKavli}
\affiliation{\UH}

\author[0000-0001-8832-4488]{Daniel Huber}
\affiliation{\UH}
\affiliation{\Sydney}

\author[0000-0003-3020-4437]{Yaguang Li}
\affiliation{\UH}

\author[0000-0003-4034-0416]{Travis S. Metcalfe}
\affiliation{\WDRC}

\author[0000-0001-5222-4661]{Timothy R. Bedding}
\affiliation{\Sydney}

\author[0000-0001-7664-648X]{Joel Ong}
\affiliation{\UH}

\author[0000-0003-1125-2564]{Ashley Chontos}
\affiliation{\Princeton}
\affiliation{\UH}

\author[0000-0003-3856-3143]{Ryan Rubenzahl}
\affiliation{\Caltech}

\author[0000-0003-1312-9391]{Samuel Halverson}
\affil{\JPL}

\author[0000-0002-8854-3776]{Rafael A. Garc\'ia}
\affiliation{\CEA}

\author[0000-0002-9037-0018]{Hans Kjeldsen}
\affiliation{\SAC}

\author[0000-0002-4879-3519]{Dennis Stello}
\affiliation{\Sydney}
\affiliation{\UNSW}
\affiliation{\ASTROTHREED}

\author[0000-0003-3244-5357]{Daniel R. Hey}
\affiliation{\UH}

\author[0000-0002-4588-5389]{Tiago Campante}
\affiliation{\PortoIAC}
\affiliation{\Porto}

\author[0000-0001-8638-0320]{Andrew W. Howard}
\affiliation{\Caltech}

\author[0009-0004-4454-6053]{Steven R.\ Gibson}
\affiliation{\COO}

\author{Kodi Rider}
\affiliation{\SSL}

\author[0000-0001-8127-5775]{Arpita Roy}
\affiliation{\Schmidt}

\author[0000-0002-6525-7013]{Ashley D.\ Baker}
\affiliation{\COO}

\author[0009-0002-2419-8819]{Jerry Edelstein}
\affiliation{\SSL}

\author{Chris Smith}
\affiliation{\SSL}

\author[0000-0003-3504-5316]{Benjamin J.\ Fulton}
\affiliation{\Caltech}

\author[0000-0002-6092-8295]{Josh Walawender}
\affiliation{\WMKO}

\author[0009-0008-9808-0411]{Max Brodheim}
\affiliation{\WMKO}

\author{Matt Brown}
\affiliation{\WMKO}

\author{Dwight Chan}
\affiliation{\WMKO}

\author[0000-0002-8958-0683]{Fei Dai}
\affiliation{\UH}

\author[0009-0000-3624-1330]{William Deich}
\affil{\UCO}

\author{Colby Gottschalk}
\affiliation{\WMKO}

\author{Jason Grillo}
\affiliation{\SSL}

\author{Dave Hale}
\affiliation{\COO}

\author[0000-0002-7648-9119]{Grant M.\ Hill}
\affiliation{\WMKO}

\author[0000-0002-6153-3076]{Bradford Holden}
\affil{\UCO}

\author[0000-0002-5812-3236]{Aaron Householder}
\affiliation{\MITEAPS}
\affiliation{\MITKavli}

\author[0000-0002-0531-1073]{Howard Isaacson}
\affiliation{\UCB}

\author[0000-0001-7572-5231]{Yuzo Ishikawa}
\affiliation{\JHU}

\author[0009-0006-2900-0401]{Sharon R.\ Jelinsky}
\affiliation{\SSL}

\author[0000-0001-8414-8771]{Marc Kassis}
\affiliation{\WMKO}

\author{Stephen Kaye}
\affiliation{\COO}

\author[0000-0003-2451-5482]{Russ Laher}
\affiliation{\nexsci}

\author[0009-0004-0592-1850]{Kyle Lanclos}
\affiliation{\WMKO}

\author[0000-0003-1700-5740]{Chien-Hsiu Lee}
\affiliation{\WMKO}

\author{Scott Lilley}
\affiliation{\WMKO}

\author{Ben McCarney}
\affiliation{\WMKO}

\author[0009-0005-2445-7544]{Timothy N. Miller}
\affiliation{\SSL}

\author[0009-0008-4293-0341]{Joel Payne}
\affiliation{\WMKO}

\author[0000-0003-0967-2893]{Erik A. Petigura}
\affiliation{\UCLA}

\author[0000-0003-0512-5489]{Claire Poppett}
\affiliation{\SSL}

\author{Michael Raffanti}
\affiliation{\SSL}

\author[0000-0002-6667-7028]{Constance Rockosi}
\affiliation{\UCO}

\author{Dale Sanford}
\affiliation{\UCO}

\author[0000-0002-4046-987X]{Christian Schwab}
\affil{\Macquarie}

\author[0000-0003-3133-6837]{Abby P.\ Shaum}
\affiliation{\Caltech}

\author[0009-0007-8555-8060]{Martin M.\ Sirk}
\affiliation{\SSL}

\author[0000-0001-7062-9726]{Roger Smith}
\affiliation{\COO}

\author{Jim Thorne}
\affiliation{\WMKO}

\author{John Valliant}
\affiliation{\WMKO}

\author[0009-0003-6534-0428]{Adam Vandenberg}
\affiliation{\WMKO}

\author{Shin Ywan Wang}
\affiliation{\nexsci}

\author{Edward Wishnow}
\affiliation{\SSL}

\author{Truman Wold}
\affiliation{\WMKO}

\author{Sherry Yeh}
\affiliation{\WMKO}

\author[0000-0002-6525-7013]{Ashley Baker}
\affiliation{\COO}

\author[0000-0002-6163-3472]{Sarbani Basu}
\affil{\Yale}

\author[0000-0001-9907-7742]{Megan Bedell}
\affil{\FlatironCCA}




\author[0000-0001-8934-7315]{Heather M.\ Cegla}
\affiliation{\Warwick}

\author{Ian Crossfield}
\affiliation{\Kansas}

\author{Courtney Dressing}
\affiliation{\UCB}

\author[0000-0002-9332-2011]{Xavier Dumusque}
\affiliation{\Geneva}


\author[0000-0002-5375-4725]{Heather Knutson}
\affiliation{\Caltech}

\author[0000-0002-8895-4735]{Dimitri Mawet}
\affiliation{\Caltech}

\author[0000-0002-7893-1054]{John O'Meara}
\affiliation{\WMKO}



\author[0000-0001-7409-5688]{Gu{\dh}mundur Stef{\'a}nsson}
\affil{\Amsterdam}

\author[0009-0008-2801-5040]{Johanna Teske}
\affil{\Carnegie}

\author[0000-0002-1871-6264]{Gautam Vasisht}
\affil{\JPL}

\author[0000-0002-6937-9034]{Sharon Xuesong Wang}
\affil{\Tsinghua}

\author[0000-0002-3725-3058]{Lauren M. Weiss}
\affiliation{\ND}

\author[0000-0002-4265-047X]{Joshua N. Winn}
\affil{\Princeton}

\author[0000-0001-6160-5888]{Jason T.\ Wright}
\affil{\PSUAA}
\affil{\PSUCEHW}

\correspondingauthor{Marc Hon}
\email{mtyhon@mit.edu}

\begin{abstract}
Asteroseismology of dwarf stars cooler than the Sun is very challenging due to the low amplitudes and rapid timescales of oscillations.
Here, we present the asteroseismic detection of solar-like oscillations at 4-minute timescales ($\nu_{\mathrm{max}}\sim4300\mu$Hz) in the nearby K-dwarf $\sigma$ Draconis using extreme precision Doppler velocity observations from the Keck Planet Finder and 20-second cadence photometry from NASA's Transiting Exoplanet Survey Satellite. The star is the coolest dwarf star to date with both velocity and luminosity observations of solar-like oscillations, having amplitudes of $5.9\pm0.8\,$cm$\,\text{s}^{-1}$ and $0.8\pm0.2$\,ppm, respectively. These measured values are in excellent agreement with established luminosity-velocity amplitude relations for oscillations and provide further evidence that mode amplitudes for stars with $T_{\mathrm{eff}}<\,5500\,$K diminish in scale following a $(L/M)^{1.5}$ relation. By modeling the star's oscillation frequencies from photometric data, we measure an asteroseismic age of $4.5\pm0.9\,\rm{(ran)} \pm 1.2\,\rm{(sys)}$\,Gyr. 
The observations demonstrate the capability of next-generation spectrographs and precise space-based photometry to extend observational asteroseismology to nearby cool dwarfs, which are benchmarks for stellar astrophysics and prime targets for directly imaging planets using future space-based telescopes. 
\end{abstract}

\section{Introduction} \label{sec:intro}

\begin{figure}
    \centering
	\includegraphics[width=1.\columnwidth]{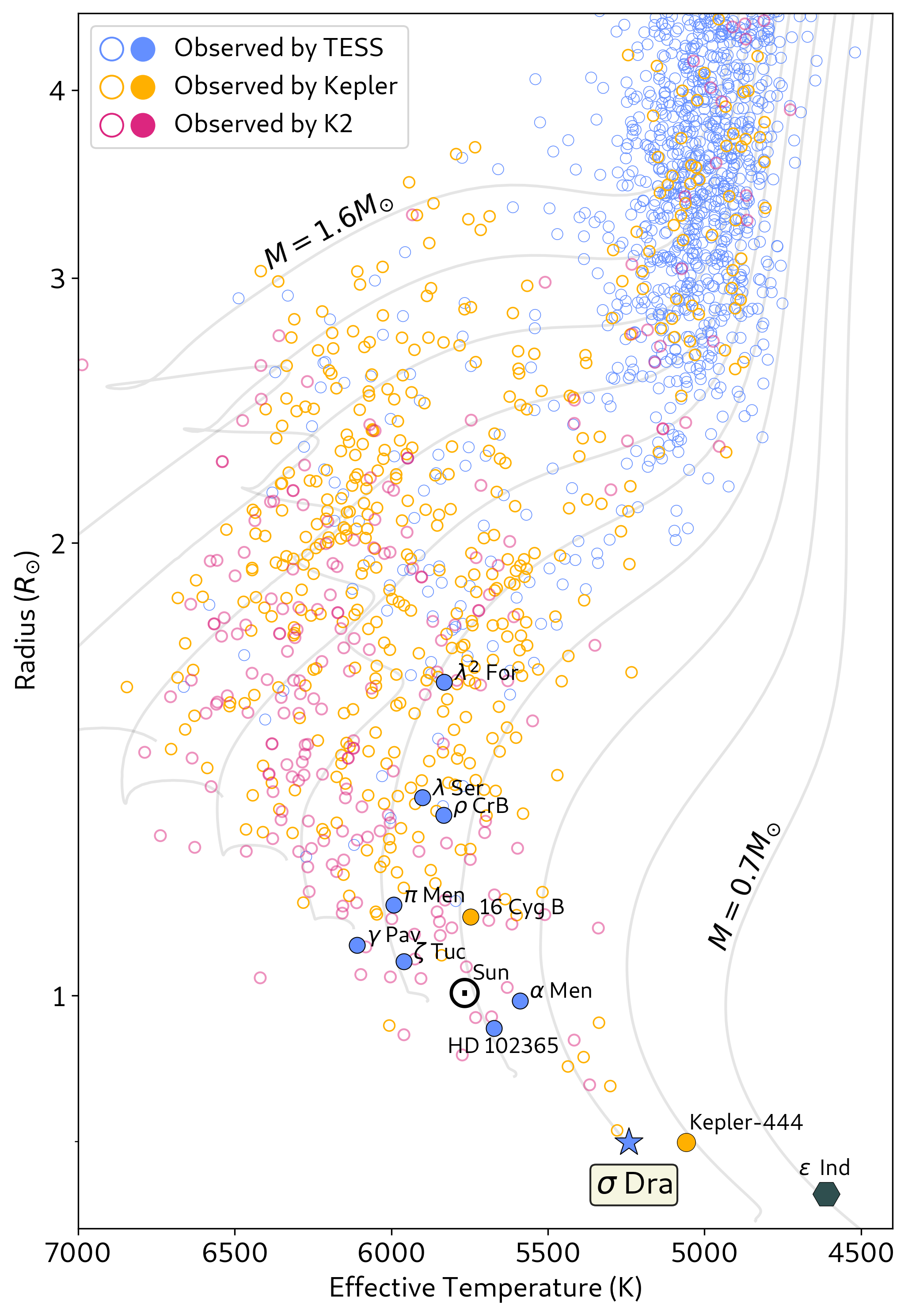}
    \caption{Radius versus effective temperature of low-mass dwarf and subgiant stars with detected solar-like oscillations. $\sigma$ Dra is the coolest dwarf star to have solar-like oscillations detected using TESS photometry to date. Only several cool dwarfs currently have solar-like oscillations detected from photometry, including Kepler-444 \citep{Campante_2015}. Also shown is $\epsilon$ Ind, the coolest star with detected solar-like oscillations, but only from radial velocity observations \citep{Campante_2024}.
    Also shown are other dwarfs and subgiants with solar-like oscillations observed by TESS \citep{Hatt_2023}, by the \textit{Kepler} mission \citep{Lund_2017, Serenelli_2017, SilvaAguirre_2017}, and by the \textit{K2} mission \citep{Chaplin_2015, Lund_2016, Gonzalez_2023, Lund_2024}. Highlighted are several dwarf stars for which solar-like oscillations have been detected and analyzed using space-based photometry  \textemdash{} $\alpha$ Men \citep{Chontos_2021}, 16 Cyg A/B \citep{Metcalfe_2012}, $\lambda^2$ For \citep{Nielsen_2020}, $\pi$ Men, $\gamma$ Pav, $\zeta$ Tuc \citep{Huber_2022}, $\rho$ CrB \citep{Metcalfe_2021}, $\lambda$ Ser \citep{Metcalfe_2023}, and HD 102365. 
    Radii and effective temperatures in the plot are adopted from the TESS Input Catalog (\citealt{TIC}, DOI: \dataset[10.17909/fwdt-2x66]{http://dx.doi.org/10.17909/fwdt-2x66}), with MIST solar-metallicity evolutionary tracks \citep{Choi_2016} displayed in the background.}
    \label{fig:fighr}
\end{figure}

Asteroseismology provides a powerful tool for measuring the fundamental properties of Sun-like stars, including masses and ages. However, the oscillation amplitudes of main-sequence stars cooler than the Sun are small, which makes detecting such oscillations difficult even with high-precision space-based photometry. A solution is to perform asteroseismology using radial velocities (RV), which are much less affected by stellar granulation noise than photometry \citep{harvey_techniques_1988, Grundahl2007, kjeldsen_amplitudes_2011} and permit a higher signal-to-noise detection of stellar oscillations. Indeed, the majority of asteroseismic detections before the era of space-based photometry were made using RVs \citep[see reviews by][]{Chaplin_2013, bedding_solar-like_2014, Garcia_2019}, including several Sun-like stars such as $\alpha$ Cen A \& B \citep{carrier_asteroseismology_2006,Kjeldsen_2005}, 18 Sco \citep{Bazot_2011}, $\tau$ Ceti \citep{teixeira_solar-like_2009}, and $\mu$ Ara \citep{bouchy_p-mode_2001}. Even so, the primary targets were nearby stars with temperatures similar to or hotter than the Sun, as existing instruments required many nights of observation to detect Doppler variations at the cm$\,\text{s}^{-1}$ level. Subsequently, NASA's \textit{Kepler} mission provided exquisite photometry of several main sequence stars cooler than the Sun \citep{Campante_2015, Lund_2017}, but their faintness is prohibitive for detailed ground-based follow-up. 

Two recent developments have enabled a more systematic application of asteroseismology to nearby cool dwarfs. First, the introduction of 20-second cadence photometry in NASA's Transiting Exoplanet Survey Satellite's (TESS, \citealt{Ricker_2014}) extended mission has greatly improved the photometric precision of bright stars \citep{Huber_2022}. Second, next-generation RV spectrographs on 8$\,$m class-telescopes like the Echelle SPectrograph for Rocky Exoplanets and Stable Spectroscopic Observations (ESPRESSO, \citealt{Pepe_2021}) aboard the Very Large Telescope (VLT) now have the efficiency to obtain high-cadence RVs with extreme precision. A prime example is the recent detection of oscillations in $\epsilon$\,Ind~A with VLT/ESPRESSO, making it the coolest dwarf with detected oscillations to date (\citealt{Campante_2024}; see also \citealt{Lundkvist_2024}).


Here, we present the clear detection of solar-like oscillations in the nearby K-dwarf $\sigma$ Dra (Alsafi; HIP 96100; HD 185144) from TESS 20-second cadence photometry and from radial velocity data by the Keck Planet Finder (KPF). Observations of $\sigma$ Dra are the first asteroseismic observations with KPF, which is a newly-commissioned RV spectrograph at the W. M. Keck Observatory in Hawai`i. The dedicated fast-readout mode of KPF (readout time 15 seconds), combined with the 10-m aperture of the Keck telescope, makes KPF a highly efficient RV asteroseismology instrument.

To date, $\sigma$ Dra is the coolest dwarf star that has oscillations detected by TESS (Fig. \ref{fig:fighr}), which also makes it the coolest dwarf star
for which both luminosity and velocity measurements of its oscillations has been obtained. Understanding the relationship between luminosity and velocity measurements of solar-like oscillations is useful for improving models of stellar convection (e.g., \citealt{Houdek_1999, Houdek_2010, Zhou_2021}) and for estimating activity levels of exoplanet hosts \citep{yu_predicting_2018}. 
$\sigma$ Dra is a particularly valuable target for examining and calibrating this relationship, given the scarcity of measured solar-like oscillations in dwarf stars cooler than the Sun.

At a distance of 5.76 parsecs from the Sun, $\sigma$ Dra is among the closest main-sequence stars to the Solar System \citep{Reyle_2021} and is an important benchmark target in stellar astrophysics. It serves as a standard star for radial velocity measurements \citep{Soubiran_2018} and for the calibration of photometric systems (e.g., \citealt{Salsi_2020, Huang_2015, Bell_1994}). In addition, the star is a prime target for long-term stellar activity monitoring (e.g., \citealt{Boro_2018, Isaacson_2010, Martinez_2010, Wilson_1963}) and detailed chemical abundance surveys in the solar neighborhood (e.g., \citealt{Ramirez_2012, Taut_2020}). 
Due to its proximity, $\sigma$ Dra has also been a target of interest in searches for nearby exoplanets (e.g., \citealt{Beichman_1999, Wittenmyer_2006, Fischer_2014, Motalebi_2015, Rosenthal_2021}). Despite the lack of confirmed planets orbiting $\sigma$ Dra to date, it remains a promising candidate host star for exoplanet surveys using the latest radial velocity instruments \citep{Gibson_2016, Jurgenson_2016} and direct imaging approaches \citep{Bowens_2021, Werber_2023}. 

$\sigma$ Dra is identified as a target of high priority (Rank A) in the NASA ExEP Mission Star List for the Habitable Worlds Observatory (HWO, \citealt{Mamajek_Karl}). Therefore, the detailed characterization of the star's fundamental properties\textemdash{}such as its luminosity and age\textemdash{}is critical for pinpointing the extent of its habitable zone (e.g., \citealt{Kopparapu_2013, Rushby_2013, Kane_2014, Kane_2018}), assessing the long-term dynamical stability of planetary orbits \citep{Davies_2014b}, and assessing the habitability timescales of planetary atmospheres (e.g., \citealt{Lammer_2018, Bixel_2020}). With the detailed profiling of photometric, spectroscopic, and activity measurements for HWO targets well underway \citep{Harada_2024}, asteroseismology from TESS and next-generation extreme-precision radial velocity (EPRV) instruments is uniquely positioned to contribute to the ambitious goal of studying $\sim\,$25 potentially habitable worlds, the goal envisioned by the Astro 2020 Decadal Survey \citep{Astro2020}.

\begin{figure}
    \centering
	\includegraphics[width=1.\columnwidth]{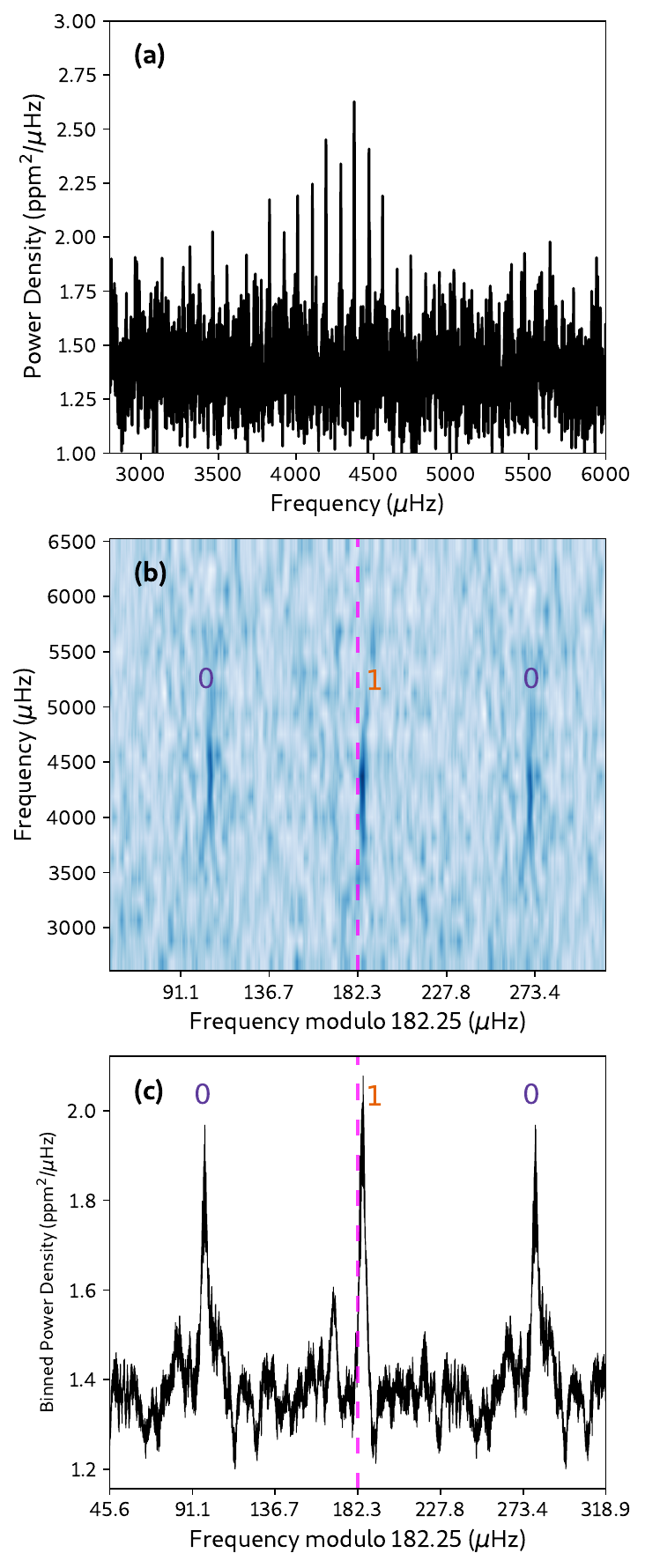}
    \caption{Solar-like oscillations of $\sigma$ Dra as observed from 14 Sectors of TESS photometry (July 2021 to January 2023). (a) The power spectral density of the star smoothed with 3$\,\mu$Hz-wide boxcar filter. (b) The replicated \'echelle diagram of the smoothed power spectrum \citep{Bedding_2012_replicated_echelles}. (c) The collapsed replicated \'echelle diagram. The dashed line in panels (b) and (c) corresponds to $\Delta\nu\,$=$\,$182.25\,$\mu$Hz and the numbers identify the $l=0$ and $l=1$ mode sequences.}
    \label{fig:fig1}
\end{figure}

\section{TESS 20-Second Cadence Photometry} \label{section:tess_phot}
\begin{figure*}
    \centering
	\includegraphics[width=2.\columnwidth]{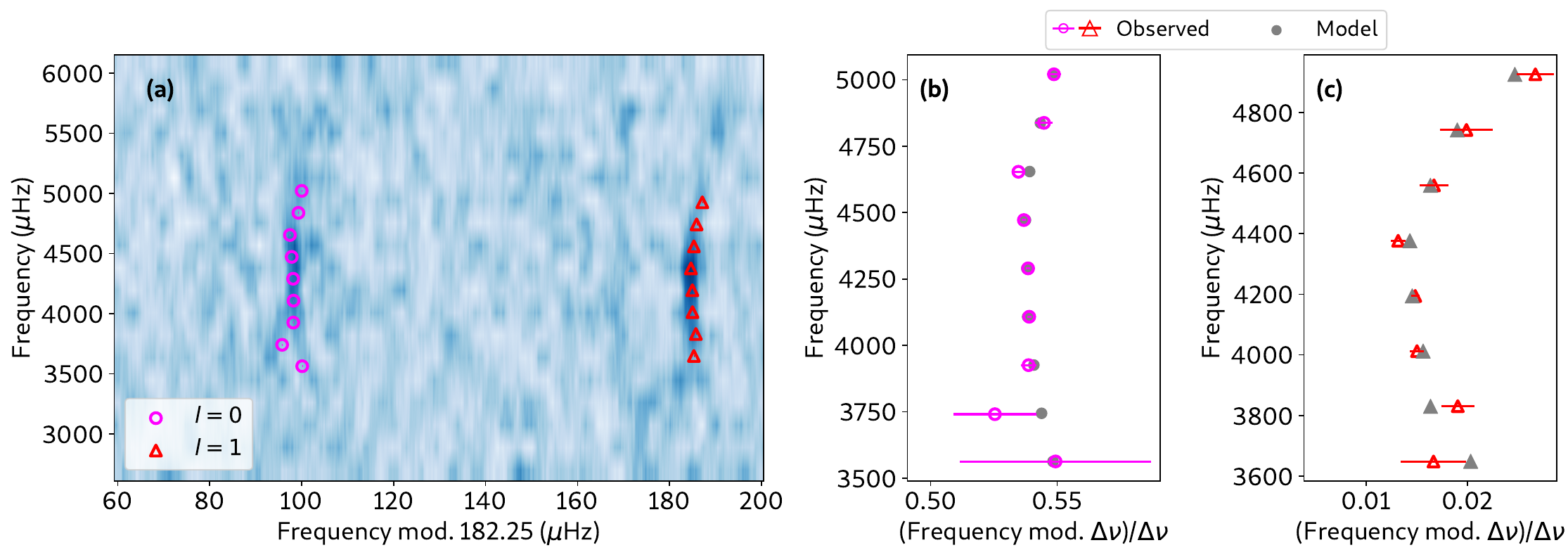}
    \caption{Mode identification of $\sigma$ Dra's oscillation spectrum from TESS photometry. (a) Extracted radial ($l=0$) and dipole ($l=1$) oscillation mode frequencies plotted on the \'echelle diagram of the smoothed power spectrum.  (b,c) Comparison of observed mode frequencies with those predicted by the best-fitting stellar model by the TM2 modeling team (Section \ref{section:stellar_modeling}), which returned a reduced $\chi^2$ value of 1.02.}
    \label{fig:fig3}
\end{figure*}

Located within TESS's Northern Continuous Viewing Zone, $\sigma$ Dra was observed at 20-second cadence during 14 sectors between July 2021 and January 2023, corresponding to Sectors 41--60. The use of 20-second cadence data places the high-frequency pulsations of the star well below the Nyquist frequency of 25 mHz and diminishes the attenuation of oscillation amplitudes due to averaging during the exposures. Importantly, TESS's 20-second cadence data also mitigates the reduction in effective exposure time as a result of onboard cosmic ray mitigation, which significantly
improves the photometric precision of observations for bright stars ($V\apprle8$), subsequently increasing the detectability of oscillation modes  \citep{Huber_2022}. Here, we use science-ready PDCSAP light curves as reduced by the TESS Science Processing Operations Center (SPOC) pipeline \citep{Jenkins_2016}, which are available on the Mikulski Archive for Space Telescopes (\citealt{TESS_LC}, DOI: \dataset[10.17909/t9-st5g-3177]{http://dx.doi.org/10.17909/t9-st5g-3177}). The PDCSAP light curve from each sector was median-normalized and subjected to a 5-$\sigma$ outlier clipping before concatenation to form the full light curve for our analysis.

The power spectrum of $\sigma$ Dra is shown in Fig. \ref{fig:fig1}a-b, where we find excess power corresponding to solar-like oscillations at frequencies of 3500--5000\,$\mu$Hz. The photometric data has a white noise level of 0.2 ppm based on the noise floor estimates at frequencies above $7000\,\mu$Hz from the amplitude spectrum. Using the \texttt{pySYD} pipeline \citep{chontos_pysyd_2022}, we measured $\nu_{\mathrm{max}} = 4250\pm150\,\mu$Hz and a large frequency separation of $\Delta\nu=182.3\pm0.3\,\mu$Hz. The revised Asteroseismic Target List \citep{Hey_2024} predicts a detection likelihood of 1, albeit at $\nu_{\mathrm{max}}$ of $3762\,\mu$Hz. The discrepancy between the predicted and the observed $\nu_{\mathrm{max}}$ values is due to the Asteroseismic Target List adopting a \textit{Gaia} GSP-Phot \citep{Andrae_2023} $\log (g)$ value of 4.5007 dex for its prediction using a scaling relation for $\nu_{\mathrm{max}}$ (Eqn. 19 of \citealt{Hey_2024}). This adopted $\log (g)$ is significantly lower than the seismically determined value of 4.589 dex presented in Section \ref{section:stellar_modeling}.

The frequency \'echelle diagram of the power spectrum (Fig. \ref{fig:fig1}b) reveals oscillation modes forming two distinct ridges. We identified the left ridge with $l=0$ modes and the other with $l=1$ modes, by relying on established relations between $\Delta\nu$ and the p-mode phase term, $\epsilon$ \citep{White_2011, Ong_2019} and by comparing the mode pattern (e.g., \citealt{Bedding_2010}) with that from Kepler-444, which is another K-type dwarf with a similar $\Delta\nu$ \citep{Campante_2015}. To measure the frequencies of these oscillation modes, we adopted `peakbagging' approaches using both Lorentzian mode-profile fitting \citep{Li_2020, Breton_2022} and iterative sine-wave fitting \citep{Frandsen_1995, Kjeldsen_2005, Bedding_2007}. Table \ref{table:mode_freq} presents the final list of mode frequencies, where all peakbagging approaches agreed within one standard deviation, adopting values from a single Lorentzian mode-profile fitting method.

Figure \ref{fig:fig3}a shows the extracted mode frequencies along with their identification as radial ($l=0$) and dipole ($l=1$) modes. The collapsed \'echelle diagram of the oscillation spectrum, summed over 10 orders, is shown in Figure \ref{fig:fig1}c. There is a hint of quadrupole ($l=2$) modes to the left of the $l=0$ peak but the presence of statistically significant modes could not be determined with confidence within the oscillation spectrum on a mode-by-mode basis.

\begin{table}
\centering
\caption{Oscillation frequencies of $\sigma$ Dra. Frequencies $f_{\mathrm{phot}}$ were extracted from TESS photometry, whereas oscillation frequencies $f_{\mathrm{vel}}$ were extracted from KPF radial velocity data. The order and spherical degree of each mode are indicated by $n$ and $l$, respectively. Corrections for line-of-sight motion \citep{Davies_2014a} have not been applied to these frequencies. Such corrections are recommended when comparing observed modes to those predicted from models due to a non-negligible shift in frequency ($\sim0.4\,\mu$Hz at 5000$\,\mu$Hz).  }\label{table:mode_freq}
\begin{tabular}{cccccc}
\toprule
$n$ & $l$ & $f_{\mathrm{phot}}$  & $\sigma_{f, \mathrm{phot}}$ & $f_{\mathrm{vel}}$  & $\sigma_{f, \mathrm{vel}}$ \\
 & & ($\mu$Hz) & ($\mu$Hz) & ($\mu$Hz) & ($\mu$Hz)\\
\midrule
17 & 1 & - & - & 3467.56 & 3.88 \\
18 & 0 & 3562.6 & 6.9 & - & - \\
18 & 1 & 3648.79 & 0.59 & - & -\\
19 & 0 & 3740.8 & 3.0 & - & - \\
19 & 1 & 3830.47 & 0.30 & 3830.26 & 3.43\\
20 & 0 & 3925.17 & 0.56 &  - & - \\
20 & 1 & 4011.97 & 0.13 & 4013.60 & 3.54\\
21 & 0 & 4107.45 & 0.15 & 4101.29 & 3.32\\
21 & 1 & 4194.175 & 0.077 & 4196.95 & 3.62\\
22 & 0 & 4289.61 & 0.14 &  - & -\\
22 & 1 & 4376.10 & 0.14 & 4380.29 & 3.86\\
23 & 0 & 4471.54 & 0.16 &  - & -\\
23 & 1 & 4558.99 & 0.26 & 4555.66 & 2.42\\
24 & 0 & 4653.39 & 0.50 & 4655.30 & 4.41\\
24 & 1 & 4741.81 & 0.48 & 4750.96 & 3.44\\
25 & 0 & 4837.47 & 0.62 &  - & -\\
25 & 1 & 4925.29 & 0.34 &  - & - \\
26 & 0 & 5020.42 & 0.50 &  - & -\\
26 & 1 &  - & - & 5101.70 & 3.04 \\
27 & 0 &  - & - & 5201.34 & 4.48 \\
27 & 1 &  - & - & 5285.04 & 3.09 \\
29 & 0 &  - & - & 5564.04 & 3.96 \\
29 & 1 &  - & - & 5663.68 & 3.34 \\
30 & 0 &  - & - & 5751.37 & 3.79 \\
31 & 1 &  - & - & 6022.39 & 5.12 \\
32 & 1 &  - & - & 6217.70 & 4.01 \\
\bottomrule
\end{tabular}
\end{table}

\section{Radial Velocities}\label{section:radial_velocities}

\begin{figure*}
    \centering
	\includegraphics[width=2.\columnwidth]{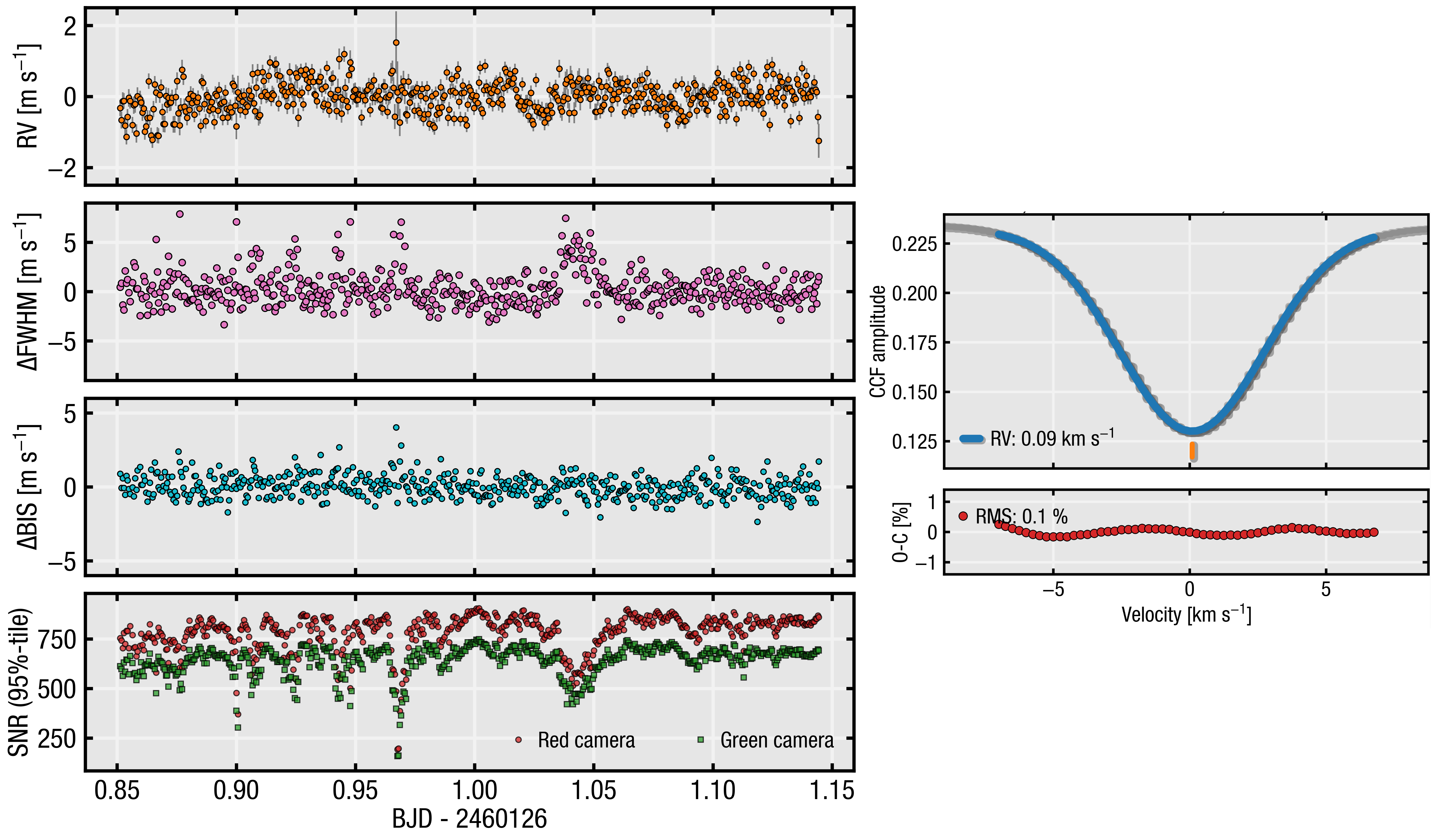}\label{figure:kpf_data}
    \caption{(Left) Radial velocity (RV) time series of $\sigma$ Dra using the Keck Planet Finder (KPF). Also presented are the median-subtracted Full Width at Half Maximum ($\Delta\text{FWHM}$) and median-subtracted Bisector Inverse Slope ($\Delta\text{BIS}$) based on measurements tabulated in Table \ref{table:kpf_data} as well as the SNR across the green and red cameras. (Right) Example of a fit to the normalized KPF cross-correlation function (CCF) of $\sigma$ Dra for a single spectrum. The CCF was computed using a binary spectral mask, as described in  \S~\ref{section:radial_velocities}.
}
    \label{fig:kpf_data}
\end{figure*}

\begin{figure*}
    \centering
	\includegraphics[width=2.05\columnwidth]{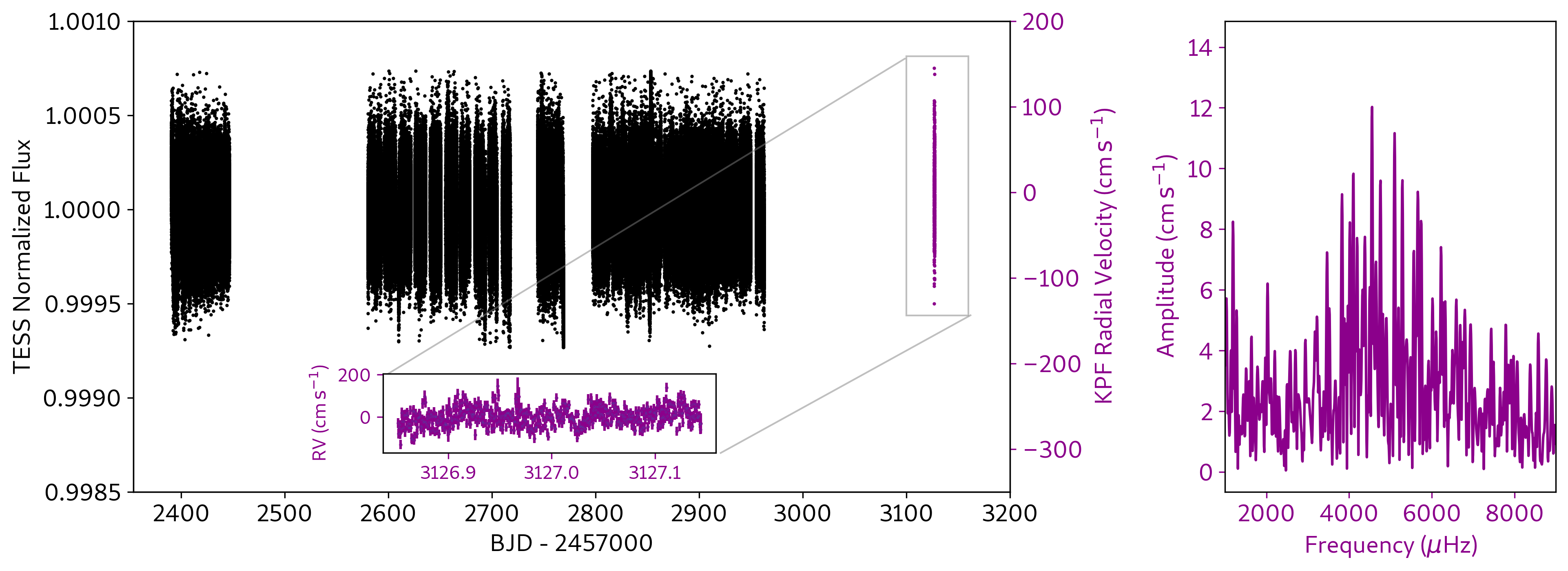}
    \caption{(Left) Observations of $\sigma$ Dra from TESS Sector 41-60 photometry in black and from Keck Planet Finder radial velocities (RVs) in purple. The inset is a zoom-in of the RV time series spanning 7 hours on June 20 2023. (Right) RV amplitude spectrum of $\sigma$ Dra, in which a power excess corresponding to solar-like oscillations can be seen at frequencies $\sim4000-6000\,\mu$Hz.}
    \label{fig:time_series}
\end{figure*}

We obtained a radial velocity (RV) time series of $\sigma$ Dra using the Keck Planet Finder spectrograph \citep[KPF,][]{gibson_kpf_2016,gibson_keck_2018,gibson_keck_2020,gibson_keck_2024}. KPF is a fiber-fed echelle spectrometer that was commissioned at the W.M. Keck Observatory in April 2023, designed to achieve an instrumental Doppler measurement precision of 30 cm s$^{-1}$ or better. The KPF main spectrometer spans $445–870$\,nm in two channels, each with three science slices, with a median resolving power of 97,000. Wavelength calibration is provided by a commercial laser frequency comb (LFC) from Menlo Systems, a broadband Fabry-Pérot etalon, and hollow cathode lamps (ThAr and UNe). KPF operates in ``standard'' and ``fast'' readout mode. The latter provides readout times of 15 seconds, designed for asteroseismic observations and other rapid-cadence applications. This rapid readout speed, combined with the 10-m aperture of Keck, makes KPF a very efficient extreme-precision radial velocity instrument for asteroseismology. While  KPF fast-readout solar observations have already demonstrated exquisite RV precision \citep{rubenzahl_staring_2023}, the observations presented here are the first KPF fast-readout observations of a star other than the Sun.

We obtained 7 hours of continuous observations of $\sigma$\,Dra with KPF on June 30, 2023 in excellent conditions. Seeing was stable at $\sim0.9"$ during the exposure sequence, with brief instances of degradation by up to $\sim0.15"$, as estimated from the guide camera images and drops in spectral signal-to-noise ratio.
We used an exposure time of 30 seconds, yielding an average SNR of 640 at $5500 \mathrm{\AA}$ (all three spectral slices combined), a median cadence of 45 seconds and a total of 560 RV datapoints. The spectra were reduced from 2D images to 1D spectra with the KPF Data Reduction Pipeline (DRP)\footnote{\url{https://github.com/Keck-DataReductionPipelines/KPF-Pipeline}}. The KPF DRP performs quadrant stitching, flat-fielding, order tracing, optimal extraction, and wavelength calibration. 

Stellar RVs are computed via a cross-correlation (CCF) mask approach using the K2 binary CCF mask used in the ESPRESSO pipeline. To measure the RVs, a Gaussian+top-hat function is used to fit an analytic mean of the CCF. A separate RV is computed for each KPF slice (3x) and camera (Green, Red). Slice RVs are combined using a weighted average, with the relative weights being proportional to photon-limited RV uncertainty in a given slice. The individual camera RVs are then median-subtracted and corrected for instrumental drift using the LFC spectra recorded before and after the stellar RV time series. The two drift-corrected stellar RV time series are then combined using a flux-weighted mean.  The KPF fast read-mode data suffers from substantial charge transfer inefficiency (CTI) in one of the four readout amplifiers on both CCDs \citep{rubenzahl_staring_2023}. To avoid adding significant systematic RV error from the contaminated spectra in these detector quadrants, spectral masks are applied to selectively mask out regions of the spectra that fell on the two bad amplifiers (one on the Green CCD, the other on the Red). A similar approach was applied to solar data by \cite{rubenzahl_staring_2023}. We note that the absolute RVs returned from the KPF CCF pipeline are consistent to within $\sim100\,$m$\,$s$^{-1}$ of the \textit{Gaia} DR3 line-of-sight velocity. Details on these data products can be found in Appendix \ref{appendix:kpf_rv}.

Figs. \ref{fig:kpf_data} and \ref{fig:time_series} show the RV observations.  The median uncertainty was 21 cm$\,\text{s}^{-1}$, which agrees well with the RMS of the time series, which was 20.3 cm$\,\text{s}^{-1}$ (although the latter also includes the signal due to oscillations). The amplitude spectrum in Fig. \ref{fig:time_series} shows a clear detection of oscillations, with an average white noise level in the amplitude spectrum of 1.89 $\text{cm}\,\text{s}^{-1}$, as measured at frequencies above 7000 $\mu$Hz. 

For completeness, we applied iterative sine-wave fitting to also extract mode frequencies from the radial velocity data. Using $\Delta\nu$ and $\epsilon$ measured from the TESS photometry in Section \ref{section:tess_phot}, we estimate the asymptotic p-mode frequencies for each radial order $n$ using $\nu=\Delta\nu\,(n+\epsilon+l/2)$, where $l$ is the mode's spherical degree. We then searched within a $\pm10\,\mu$Hz range of each asymptotic mode frequency for radial and dipole modes by identifying the frequency at which the highest power was found. A sine wave with this frequency was fitted to and subtracted from the residual time series, where their amplitudes and phases were treated as free parameters. The reported uncertainties for these modes, $\sigma_{f, {\mathrm{vel}}}$, were estimated using the analytical expression for one-sigma errors in frequency when performing a least squares fit to the data with a sinusoidal signal \citep{Montgomery_1999, Kjeldsen_2021}, and are thus only lower bounds to the actual uncertainties of the frequencies $f_{\mathrm{vel}}$ of modes extracted from the RV data.
The values of $f_{\mathrm{vel}}$ and $\sigma_{f, {\mathrm{vel}}}$ are also tabulated in Table \ref{table:mode_freq}, where we only report modes with amplitudes higher than three times the noise amplitude  ($1.8\,\text{cm s}^{-1}$), which is estimated from the mean of the amplitude spectrum between 8000 to 11000~$\mu$Hz.


\section{Asteroseismic Analysis}

\subsection{Stellar Modeling}\label{section:stellar_modeling}
As inputs for the modelling of $\sigma$ Dra, we used photometric oscillation frequencies $f_{\mathrm{phot}}$ from Table \ref{table:mode_freq}, measurements of effective temperature ($T_{\mathrm{eff}}$), metallicity ([M/H]), and interferometric stellar radius ($R$). For the oscillation frequencies, we applied corrections based on the stellar line-of-sight Doppler velocity shift \citep{Davies_2014a} assuming a radial velocity of $v_r=26.55\pm0.13\,$km$\,\text{s}^{-1}$ as reported by \textit{Gaia} DR3 \citep{Katz_2023}. The adopted values of $T_{\mathrm{eff}}=5290\pm90\,$K and [Fe/H] = $-0.23\pm0.05$ dex were obtained as averaged measurements from the PASTEL catalogue\footnote{An updated version of the catalog \citep{Soubiran_2022} gives $T_{\mathrm{eff}}=5298\,$K and [Fe/H] = $-$0.21 dex, which remain consistent with our adopted values.} \citep{Soubiran_2016}, with uncertainties derived from the dispersion across those measurements. For a radius constraint, we adopted the linear radius measurement of $R=0.778\pm0.008\,R_{\odot}$ based on interferometric measurements by \citet{Boyajian_2008}. We note that the interferometrically-derived $T_{\mathrm{eff}}$ from that study, $5299\pm32\,$K, is consistent with our value adopted from the PASTEL catalog.

To derive stellar properties for $\sigma$ Dra, modelling runs were conducted by four teams (TM1, TM2, YL, JO).

\subsubsection{Modeling by TM1} \label{section:tm1_modeling}
\
The modeling approach by TM1 used version 1.3 of the Asteroseismic Modeling Portal \citep[AMP;][]{Metcalfe_2009, creevey_characterizing_2017}, which uses the Aarhus stellar evolution code \citep[ASTEC;][]{ASTEC} and adiabatic pulsation code \citep[ADIPLS;][]{ADIPLS} to interface with a parallel genetic algorithm \citep{Charbonneau_1995, Metcalfe_2003}.
The stellar models adopt the OPAL 2005 equation of state \citep{Iglesias_1996}, OPAL opacities \citep{Rogers_2002} supplemented with low-temperature values from \citet{Ferguson_2005} at low temperatures, the \citet{Grevesse_1998} solar mixture, NACRE nuclear reaction rates \citep{Angulo_1999}, a gray atmosphere with Eddington $\text{T}-\tau$ integration \citep{Eddington_1926}, the \citet{BV_1958} mixing-length formalism, diffusion and settling of helium (but not heavy elements) from \citet{Michaud_1993}, and the \citet{Kjeldsen_2008} prescription of the asteroseismic surface correction. No convective overshooting, rotation, or magnetic fields are included in the models.

The model-fitting pipeline optimizes stellar mass ($M$) from 0.75 to 1.75 $M_\odot$, initial metal fraction ($Z$) from 0.008 to 0.05 (equally spaced in $\log Z$), initial helium mass fraction ($Y_{\rm i}$) from 0.22 to 0.32, and the mixing-length parameter ($\alpha$) from 1.0 to 3.0. The stellar age is optimized internally during each model evaluation by matching the the observed average frequency spacing of
consecutive $l=0$ modes, which decreases almost monotonically from the zero age main-sequence to the base of the red giant branch \citep{Christensen-Dalsgaard_1993}. The underlying genetic algorithm uses two-digit decimal encoding, such that there are 100 possible values for each varied parameter. The values and uncertainties of the asteroseismic properties were determined from the likelihood-weighted mean and standard deviation of all models sampled by the genetic algorithm.

\subsubsection{Modeling by TM2}
The modeling approach by TM2 used AMP version 2.0 \citep{Metcalfe_2023}, which now interfaces the AMP genetic algorithm with the Modules for Experiments in Stellar Astrophysics \citep[MESA;][]{paxton_modules_2011_, paxton_modules_2013_, paxton_modules_2015_, paxton_modules_2018_, paxton_modules_2019_} version \texttt{r12778} stellar evolution code and the GYRE adiabatic pulsation code version 6.0 \citep{townsend_gyre_2013_}. While otherwise similar to AMP v1.3, this AMP version uses MESA's default equation of state, which is primarily from OPAL \citep{Rogers_2002} and SVCH \citep{Saumon_1995}. In addition, this version adopts the \citet{Cox_1968} mixing length formalism, diffusion and settling of helium and heavy elements from \citep{Thoul_2004}, and the two-term \citet{Ball_2014} prescription of the asteroseismic surface correction. No convective overshooting, rotation, or magnetic fields are included in the models. Model-fitting was performed using a similar approach to that described in Section \ref{section:tm1_modeling}.

\subsubsection{Modeling by YL} \label{section:yl_modeling}
The modeling approach by YL used version \texttt{r15140} of MESA alongside the GYRE adiabatic pulsation code version 6.0. The stellar models similarly adopt MESA's default equation of state. The opacity tables are selected from a combination of electron conduction opacities \citep{Cassisi_2007}, OPAL radiative opacties \citep{Iglesias_1993, Iglesias_1996}, low-temperature opacities \citep{Ferguson_2005} and data at the high-temperature Compton-scattering regime \citep{Buchler_1976}. The models use the \citet{Asplund_2009} solar mixture, JINA REACLIB nuclear reaction rates \citep{Cyburt_2010}, a gray atmosphere with Eddington $\text{T}-\tau$ integration \citep{Eddington_1926}, and the \citet{Henyey_1965} formalism for convection, whose efficiency is controlled by a mixing length parameter $\alpha$. The \citet{Ball_2014} two-term correction is used as the prescription for the asteroseismic surface effect. Effects of atomic diffusion, gravitational settling, and overshooting are not included in the models.

The model search was performed across a grid having free parameters $M\in[0.7, 2.3]\,M_{\odot}$, $Y_{\rm i}\in[0.20, 0.33]$, $\mathrm{[M/H]}\in[-0.94, 0.56]$, and $\alpha\in[1.30, 2.70]$, with all parameters uniformly sampled in a quasi-random Sobol sequence of length 8,191.
The fitting utilized a maximum likelihood approach involving the $\chi^2$ minimization of fits to both classical (non-seismic) and seismic (mode frequencies) constraints.The seismic $\chi^2$, in particular, is the sum of the residuals of the fits across all oscillation modes weighted by their observational uncertainties.

\subsubsection{Modeling by JO}

The modeling approach by JO used MESA version \texttt{r12778} and version 5.2 of the GYRE adiabatic pulsation code. The stellar models adopt the MESA default equations of state and use the same opacity tables as in Section \ref{section:yl_modeling}. However, the opacity values are set to be varied throughout the atmosphere to be consistent with the local thermodynamic state\footnote{Set by \texttt{atm\_T\_tau\_opacity = 'varying'} in the MESA control inlist.}. The modeling also uses the \citet{Grevesse_1998} solar mixture, JINA REACLIB nuclear reaction rates \citep{Cyburt_2010}, and a gray atmosphere with Eddington $\text{T}-\tau$ integration \citep{Eddington_1926}. In addition, the models adopt the \citet{Cox_1968} mixing-length formalism, the diffusion and settling of helium and heavy elements as described by \citet{Burgers_1969}, a small amount of exponential convective overshoot ($f_{\mathrm{ov}}$ = 0.008), and a combination of the \citet{Roxburgh_2016} phase-matching method and the \citet{Ball_2018} two-term correction for the asteroseismic surface correction. Neither rotation nor magnetic fields are included in the models.

As described in \citet{Ong_2021}, model-fitting was performed across a Sobol-sampled parameter space comprising $M\in[0.7, 1.7]\,M_{\odot}$,
$Y_{\rm i} \in [0.23, 0.37]$, initial metallicity [Fe/H]$_0\in[-1.5, +0.25]\,$dex, and $\log_{10}\alpha_{\text{MLT}}\in[-0.2, +0.2]\,\alpha_{\mathrm{MLT, \odot}}$, with $\alpha_{\mathrm{MLT, \odot}}=1.82$. The log-likelihood function was the sum of equally-weighted sum $\chi^2$ statistics for the observed [Fe/H], $T_{\mathrm{eff}}$, and $R$, as well as a reduced $\chi^2$ statistic for the surface-corrected mode frequencies.
The derived fundamental properties were estimated as the marginal posterior median and $\pm1\sigma$ quantiles of weighted averages of the properties over the grid of models, with the weights being proportional to the likelihood function and inversely proportional to the assumed prior distribution of the grid.

\subsubsection{Modeling Results}
Table \ref{table:properties} reports the results across the modeling teams. We find a reasonable agreement for stellar mass and luminosity across modelling runs, with individual estimates spanning 0.81--0.85\,$M_{\odot}$ and 0.40--0.42\,$L_{\odot}$. Inferred stellar ages are significantly more dispersed, with a range of 3.2--6.1\,Gyr around a median value of $4.28$ Gyr
across runs. We adopt the self-consistent set of stellar parameters from the TM2 modeling result, which yields the smallest difference in modelled parameters to the median across runs. Our final estimate reports random uncertainties from this adopted solution and systematic uncertainties in the form of standard deviation of reported parameters across teams.

Adding random and systematic uncertainties in quadrature results in a total fractional age uncertainty of about 35\%. The large uncertainty relative to cool dwarfs modeled by \textit{Kepler} (e.g., \textit{Kepler}-444, $\sim10\%$, \citealt{Campante_2015}) and warmer dwarfs observed by TESS ($\sim20\%$, \citealt{Huber_2022}) is likely a consequence of the absence of quadrupole ($l=2$) modes. As a result, information related to the small frequency separation $\delta\nu_{02} = \nu_{n, l=0} +\nu_{n-1, l=2}$ was not present in the modeling. The quantity $\delta\nu_{02}$ is sensitive to central hydrogen abundance and is thus indicative of the main sequence lifetime \textemdash{} and thus the age \textemdash{} of the star \citep{JCD_1984}. We discuss the possibility of detecting quadrupole modes in Section \ref{sec:quad}, which would substantially improve age estimates. Nonetheless, our derived ages are consistent with the values of $3.0-5.7\,$Gyr estimated in the literature using a broad range of measurements, including lithium abundances \citep{Ramirez_2012}, activity measurements \citep{Stanford-Moore_2020}, and gyrochronology \citep{Barnes_2007, Mamajek_2008}.

\begin{table*}[ht]
    \centering
    \caption{Stellar Properties for $\sigma$ Dra. The star's magnitude is adopted from TESS Input Catalog \citep{Stassun_2019}, while its parallaxes are from \textit{Gaia} DR3 \citep{Gaia_DR3}. The [M/H] and radius values are provided by \citet{Soubiran_2016} and \citet{Boyajian_2008}, respectively.}\label{table:properties}
    \begin{tabular}{cccccc}
        \toprule
        \multicolumn{6}{c}{\textbf{General Properties}} \\ 
        \midrule
        \multicolumn{2}{c}{HIP ID} & \multicolumn{4}{c}{96100} \\
        \multicolumn{2}{c}{HD ID} & \multicolumn{4}{c}{185144} \\
        \multicolumn{2}{c}{TIC ID} &  \multicolumn{4}{c}{259237827} \\
        \multicolumn{2}{c}{TESS Magnitude} & \multicolumn{4}{c}{3.94} \\
        \multicolumn{2}{c}{Distance (pc)}  & \multicolumn{4}{c}{$5.764 \pm 0.002$} \\
        \midrule
        \multicolumn{6}{c}{\textbf{Adopted Properties for Modelling}} \\ 
        \midrule
        \multicolumn{2}{c}{$T_{\mathrm{eff}}$ (K)} & \multicolumn{4}{c}{\hspace{-9.5pt}$5290\pm90$} \\
        \multicolumn{2}{c}{$[$M/H$]$ (dex)} & \multicolumn{4}{c}{\hspace{-7.5pt}$-0.23\pm0.05$} \\
        \multicolumn{2}{c}{Radius ($R_{\odot}$)} & \multicolumn{4}{c}{$0.778\pm0.008$}\\
        \midrule
        \multicolumn{6}{c}{\textbf{Model-Derived Parameters}} \\
        \midrule
        & TM1 & TM2 & YL & JO & Adopted \\
        \midrule
        Mass ($M_{\odot}$) & 0.81 & 0.84 & 0.85 & 0.84 &  \hspace{1pt}$0.84\pm0.01\,$(ran) $\pm\,0.02\,$(sys)  \\
        Radius ($R_{\odot}$) & 0.763 & 0.772 & 0.780 & 0.772 & \hspace{1pt}$0.772\pm0.005\,$(ran) $\pm\,0.007\,$(sys)  \\
        Luminosity ($L_{\odot}$) & 0.42 & 0.42 & 0.40 & 0.42 & \hspace{1pt}$0.42\pm0.04\,$(ran) $\pm\,0.01\,$(sys)  \\
        $\log (g)$ (cgs) & 4.584 & 4.589 & 4.585 & 4.590 & \hspace{1pt}$4.589\pm0.008\,$(ran) $\pm\,0.003\,$(sys)  \\
        Age (Gyr) & 6.13 & 4.54 & 3.21 & 4.02 & \hspace{1pt}$4.54\pm0.92\,$(ran) $\pm\,1.23\,$(sys)  \\
        \bottomrule
    \end{tabular}
\end{table*}

\begin{figure*}
    \centering
	\includegraphics[width=2.\columnwidth]{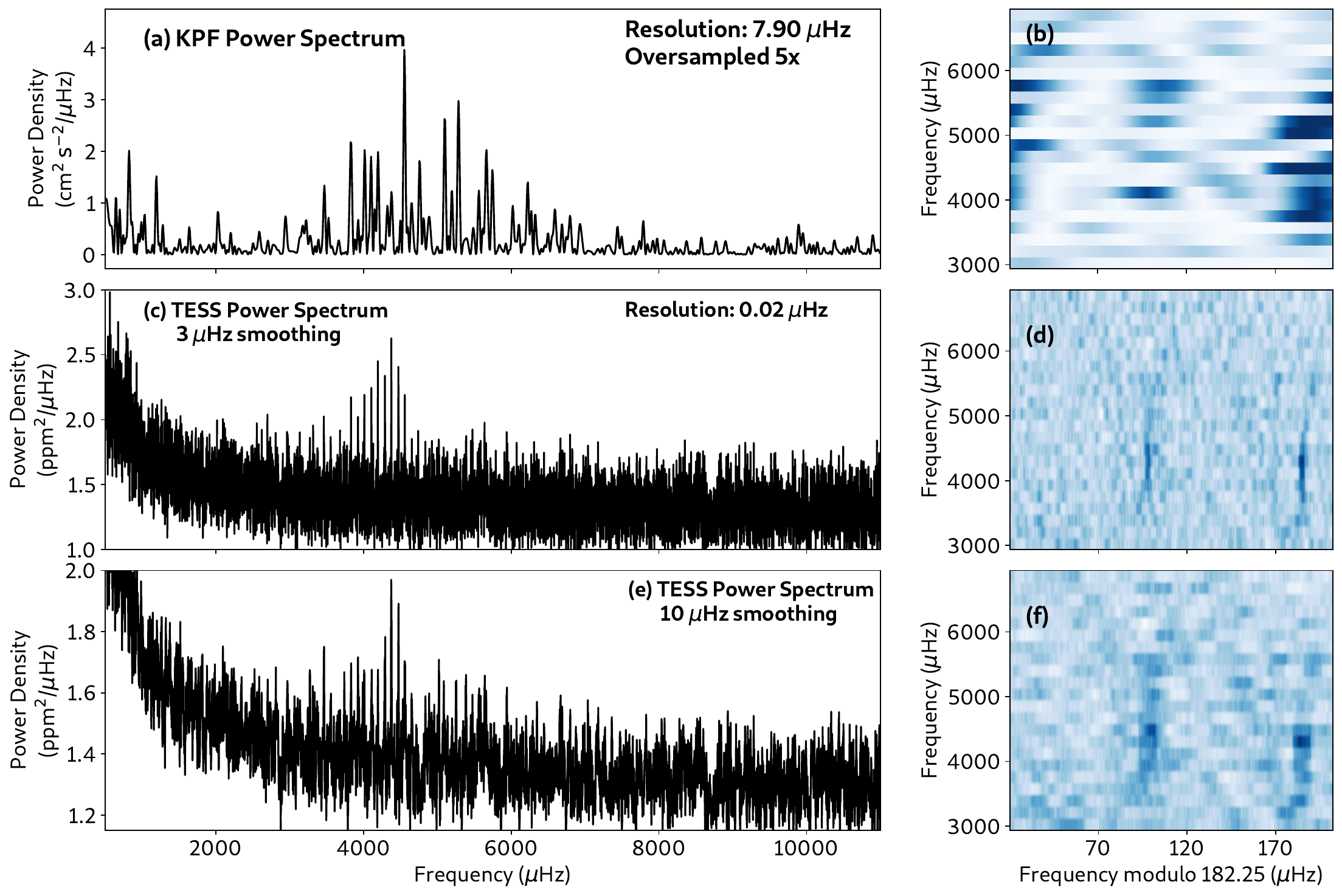}
    \caption{Oscillations of $\sigma$ Dra as observed using Keck Planet Finder (KPF) radial velocity measurements (a-b) in comparison with observations using TESS photometry (c-f). The left column presents oscillation power spectra, whereas the right column presents their corresponding frequency \'echelle diagrams.}
    \label{fig:fig_phot_vs_rv}
\end{figure*}

\subsection{Oscillations in Photometry versus Radial Velocities}
A comparison of RV and photometric power spectra is presented in Figure \ref{fig:fig_phot_vs_rv}. For low-amplitude oscillations in K-dwarfs, the lack of significant background power from granulation in the RV data is highly advantageous for increasing the S/N of the observed oscillations. As shown by the sloping background in Figs.  \ref{fig:fig_phot_vs_rv}c-d, the granulation background in photometry contributes a significant fraction of the total power in the spectrum. This limits the photometric detection of low-amplitude solar-like oscillations in TESS to the brightest stars \textemdash{} a limitation that will not be as severe for extreme-precision radial velocity measurements from KPF. Indeed, we find that the 
\'echelle diagram in Fig. \ref{fig:fig_phot_vs_rv}b reveals structure resembling the vertical mode ridges seen in photometry (Fig. \ref{fig:fig_phot_vs_rv}e), confirming that the observed power excess in the RV data corresponds to the star's oscillation modes.

\begin{figure}
    \centering
	\includegraphics[width=1.\columnwidth]{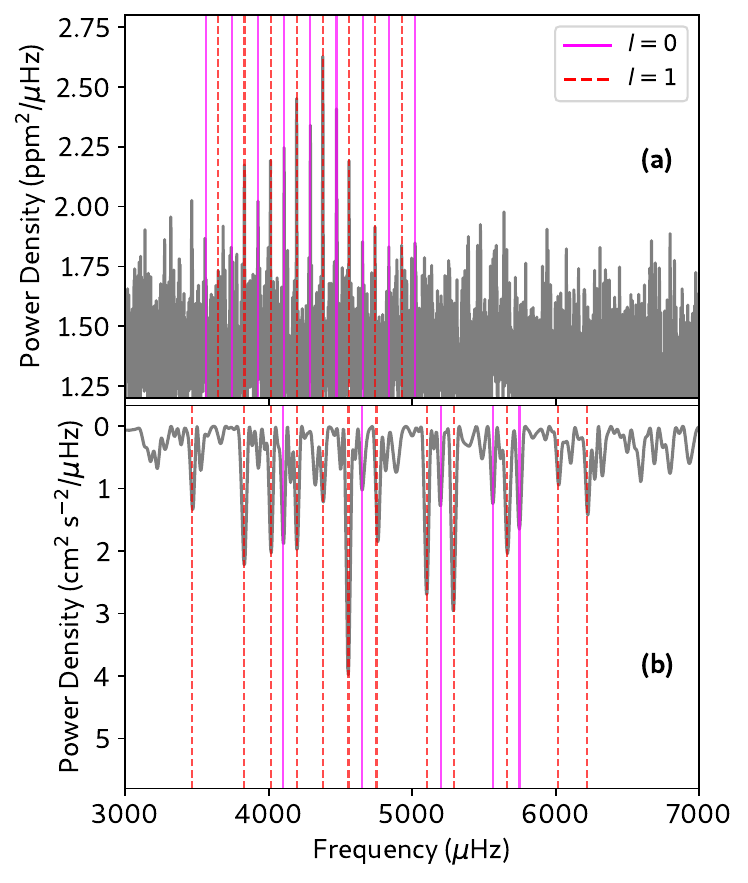}
    \caption{A comparison of identified oscillation modes (Table \ref{table:mode_freq}) from (a) TESS photometric data with those from (b) Keck Planet Finder radial velocity data, highlighting the differences in the observed width of the oscillation power excess across the two datasets.}
    \label{fig:rv_modes}
\end{figure}

There is an unusual difference in the observed width of $\sigma$ Dra's power excess across the RV and TESS photometry. While both power excesses share a low frequency limit of about 3600$\,\mu$Hz, the pre-whitening analysis of the RV data in Section \ref{section:radial_velocities} identifies mode frequencies as high as 6200$\,\mu$Hz, whereas oscillation modes are undetectable in the TESS photometry at frequencies above $5000\,\mu$Hz (see Figs. \ref{fig:rv_modes}a-d). However, Figs. \ref{fig:rv_modes}e-f show that further heavy smoothing of the TESS power spectrum reveals the marginal presence of p-modes spanning up to $6000\,\mu$Hz. The plausible reason for the presence of these modes only with heavy smoothing is the decrease of mode heights with frequency, which has been observed to occur drastically for dwarf stars cooler than the Sun \citep{Appourchaux_2014, Lund_2017}. Therefore, the heights of high-frequency modes for $\sigma$ Dra are probably below the white noise level in the TESS power spectrum. 
This effect is prominent in the TESS data because the p-modes are well-resolved from the observations, unlike the p-modes in the KPF power spectrum that are based on a time series about 1,900 times shorter than the TESS observations. 

Importantly, these differences result in a systematic offset of $\sigma$ Dra's frequency at maximum power, $\nu_{\mathrm{max}}$, between KPF RVs and TESS photometry. We therefore caution against the use of $\nu_{\mathrm{max}}$ to characterize the asteroseismic properties of solar-like oscillations in dwarf stars. This caution is further warranted by other sources of offsets in $\nu_{\mathrm{max}}$ seen within the photometric and multi-wavelength Doppler datasets of the solar oscillation spectrum \citep{Howe_2020}. The accurate characterization of solar-like oscillations is particularly relevant for RV searches of Earth-analogue planets. As detailed by \citet{Chaplin_2019}, the p-mode oscillation amplitudes of host stars can be effectively attenuated by setting exposure times
to the stars' typical oscillation timescales, identified as $1/\nu_{\mathrm{max}}$. Following this approach, an incorrect estimate of $\nu_{\mathrm{max}}$ may lead to sub-optimal observing strategies in which p-mode oscillation amplitudes are not effectively averaged out, which can significantly hinder the identification of subtle signatures within the RV data. 

\subsection{Oscillation Amplitudes and Prospects for Detections in Cooler Stars}\label{sec:amplitudes}

To measure the oscillation amplitudes from the KPF RV data, we followed the approach outlined in \citet{Kjeldsen_2005, Kjeldsen_2008}. In particular, we smoothed the RV power density spectrum heavily using a Gaussian window with a FWHM of 4$\,\Delta\nu$, followed by the fitting and subtraction of background noise. The maximum power density value corresponding to the smoothed power excess was determined, and then multiplied by $\Delta\nu/c$ for its conversion into amplitude per radial mode. Here, $c$ was taken to be $3.04$, which represents the effective number of modes per radial order that is normalised to the amplitudes of radial modes for full-disk velocity observations \citep{Kjeldsen_2008}. We measured an amplitude of $v_{\mathrm{osc}}=5.9 \pm 0.8\,$cm$\,\text{s}^{-1}$ for $\sigma$ Dra, which is approximately 33\% of that of the Sun's solar cycle-averaged value of $v_{\mathrm{osc, \odot}}=18.7\pm0.7\,$cm$\,\text{s}^{-1}$ \citep{Kjeldsen_2008}. We determined the uncertainties for $v_{\mathrm{osc}}$ by measuring the dispersion of $v_{\mathrm{osc}}$ calculated from 10,000 time series simulated to have $\sigma$ Dra's oscillations as observed with the KPF cadence and observation duration in this study. The simulations were generated using 
the asteroFLAG Artificial DataSet Generator (\texttt{AADG3}, \citealt{Ball_2018}) using mode frequencies in Table \ref{table:mode_freq} with amplitudes scaled by 
$v_{\mathrm{osc}}$, and a mode linewidth of $1.5\,\mu$Hz, a typical value based on observations of \textit{Kepler} main-sequence stars at similar temperatures to $\sigma$ Dra \citep{Lund_2017}. We additionally injected a white noise component with a level identical to that observed in the KPF time series, and thus the noise in the simulations combined contributions from shot noise, realisation noise, and the stochastic behaviour of solar-like oscillations.

Observations of $\sigma$ Dra's oscillations in both photometry and radial velocities provide the opportunity to test the \citet{Kjeldsen_1995} luminosity-velocity amplitude relation described by the following:
\begin{equation} \label{eqn:kb_95}
    A_\lambda = \frac{v_{\mathrm{osc}}/\text{cm s}^{-1}}{(\lambda/550\,\text{nm})(T_{\mathrm{eff}}/5777\,\text{K})^2}\,0.201\,\text{ppm},
\end{equation}
where $A_\lambda$ is the fractional luminosity variation due to oscillations at a given wavelength $\lambda$. At the central wavelength $\lambda=786.5\,$nm of the detector bandpass aboard TESS, we calculate $A_{786.5}$ = $1.0\,\pm\,0.1\,$ppm based on the KPF velocity amplitude. This value is in good agreement with the amplitude per radial mode of $0.8\pm0.2$ ppm estimated from TESS photometry. While a more robust test of this relation requires simultaneous luminosity and velocity measurements, $\sigma$ Dra is the solar-like oscillator with the lowest amplitude to date for which Equation \ref{eqn:kb_95} has been examined. The agreement shown here confirms the expected linearity of these low-amplitude oscillations for K-dwarfs and confirms the utility of the luminosity-velocity amplitude relation for predicting the asteroseismic detectability of cool dwarfs.

Another important relation related to asteroseismic detectability is the Doppler velocity p-mode oscillation amplitude scaling relation, which takes the form of $v_{\mathrm{osc}}\propto (L/M)^s$.  It was proposed by \citet{Kjeldsen_1995} with $s=1$, based on theoretical models presented by \citet{JCD_1983}. Subsequently, $s$ has been predicted theoretically to have a value between $0.7$ and $1.5$ (e.g., \citealt{Houdek_2002, Samadi_2007}). However, observational measurements suggest a variation with temperature \citep{Verner_2011}, for which the oscillation amplitudes of K-dwarfs diminish more rapidly than for hotter Sun-like stars with decreasing stellar luminosity. In their analysis of the K5 dwarf $\epsilon$ Ind~A, \citet{Campante_2024} presented firm evidence of this relation for a K-dwarf with $T_{\mathrm{eff}} < 5000\,$K, showing that its ${L/M}$ scaling relation can be described with an exponent ($s=1.5$) larger than for hotter dwarf stars ($s\sim0.7-1.0$). 

\begin{figure}
    \centering
	\includegraphics[width=1.\columnwidth]{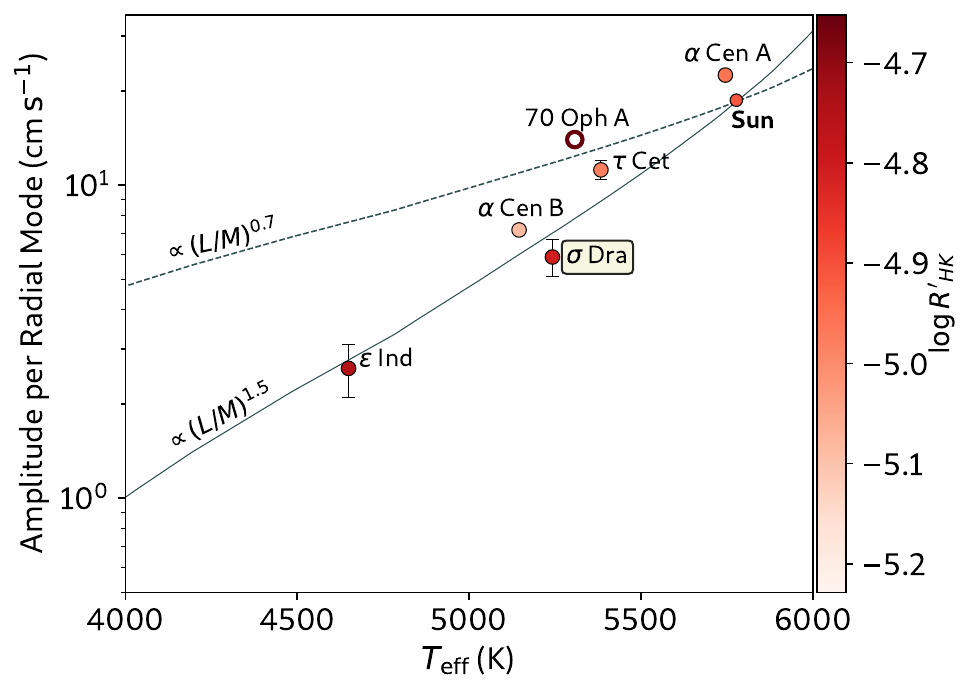}
    \caption{Comparison of the amplitude of radial mode of $\sigma$ Dra with those from other cool dwarfs with published measurements. Two scalings of the mode amplitudes are shown, differing in terms of the exponent ($s = 0.7$, dotted curve; $s = 1.5$, solid curve). The $L/M$ relations are shown using a 4.57 Gyr, solar-metallicity isochrone computed with the PAdova and TRieste Stellar Evolution Code (PARSEC, \citealt{Bressan_2012}). The value of $\log\mathcal{R}'_{HK}$ for all other stars besides $\epsilon$ Ind~A (\citealt{Mamajek_2008}) are adopted from \citet{Silva_2021}. For $T_{\mathrm{eff}}$, values for $\epsilon$ Ind~A and 70 Oph~A are from \citet{Rains_2020}, while those for other stars are from \citet{Bruntt_2010}. 
}
    \label{fig:activity_amplitude}
\end{figure}

\begin{figure}
    \centering
	\includegraphics[width=1.\columnwidth]{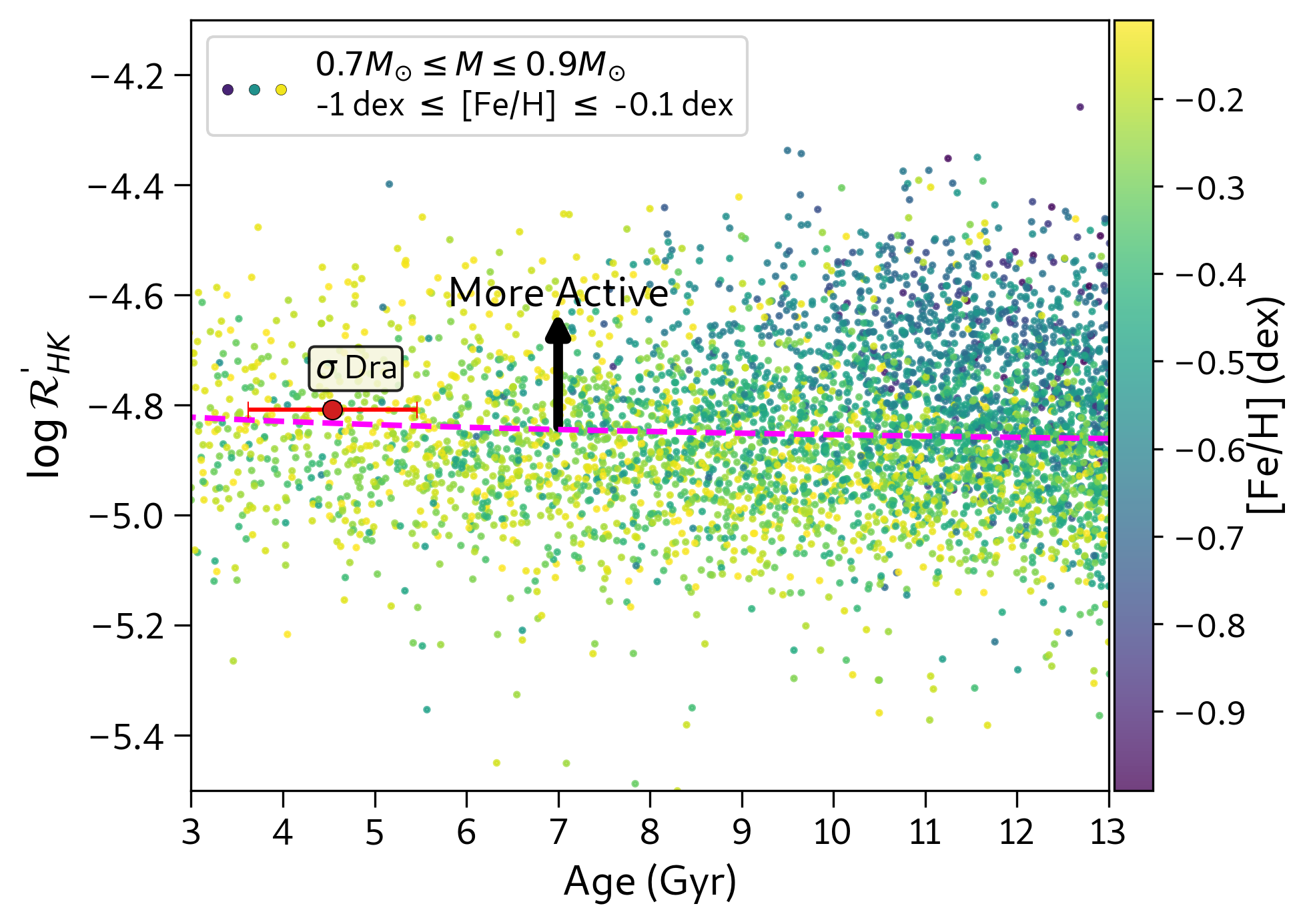}
    \caption{Age and activity of $\sigma$ Dra relative to a sample of 5,497 K-dwarfs with sub-solar metallicities from the \citet{Ye_2024} \textit{Kepler}-LAMOST catalog of FGK-dwarfs with isochrone-derived ages. The dashed line represents the adopted relation $\log\mathcal{R}_{HK}' = -0.061\log\frac{\text{Age}}{\text{Gyr}} -4.244$ that describes the age-activity relation for stars within this mass and metallicity bin. The value of $\log\mathcal{R}_{HK}'$ for $\sigma$ Dra is adopted from \citet{Boro_2018}, while its age is derived from this study. }
    \label{fig:age_rhk}
\end{figure}

In Fig. \ref{fig:activity_amplitude}, we present an update to the $v_{\mathrm{osc}}-T_{\mathrm{eff}}$ plot presented in Fig. 5 of \citet{Campante_2024}, where we now include our measurements of $\sigma$ Dra. Similar to $\epsilon$ Ind~A, the $v_{\mathrm{osc}}$ for $\sigma$ Dra favors an amplitude scaling relation with an exponent $s$ of approximately 1.5, which has been suggested by \citet{Campante_2024} to be caused by the fact that cooler stars are more likely to have higher surface magnetic activity, which suppresses observed oscillation amplitudes. Given that $\sigma$ Dra has moderate levels of chromospheric activity ($\log \mathcal{R}'_{HK} =-4.808$ dex, \citealt{Boro_2018}), these results corroborate the hypothesis that increased magnetic activity among cooler stars skews the scaling relation's exponent to larger values. This would suggest that the oscillations of old, metal-rich K-dwarfs are more likely to be detected, given their observed decrease of activity with age and metallicity (Fig. \ref{fig:age_rhk}).

\begin{figure}
    \centering
	\includegraphics[width=1.\columnwidth]{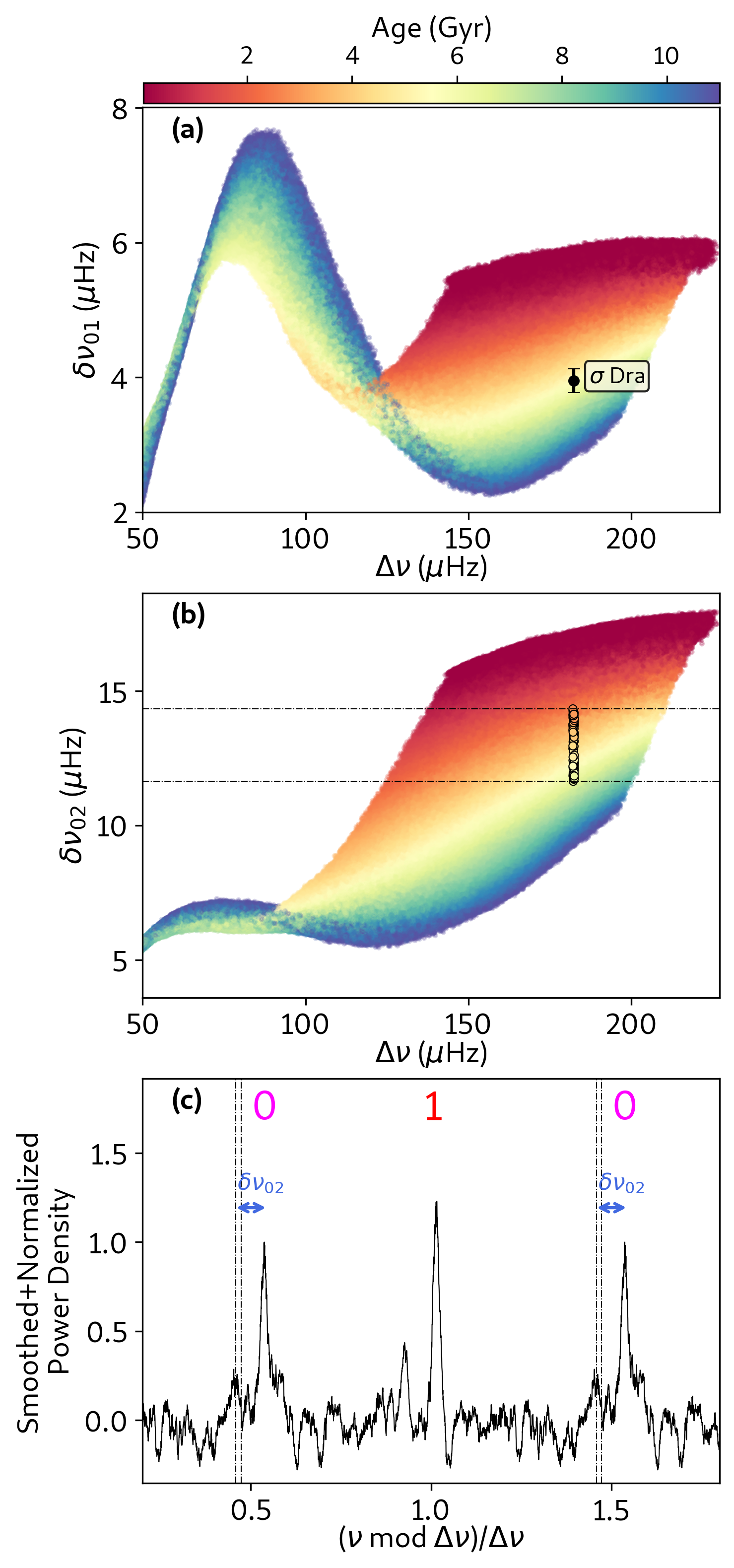}
    \caption{(a-b) Location of $\sigma$ Dra on asteroseismic diagrams relative to $M=0.7-1.0\,M_{\odot}$ models with a metallicity within 2 standard deviations of $\sigma$ Dra's [Fe/H] at $Y_i=0.277$ and $\alpha=1.9$. The grid of models are emulated using \texttt{modelflows} \citep{Hon_2024}. Panel (a) shows the position of $\sigma$ Dra on the $\delta\nu_{01}-\Delta\nu$ diagram. Given $\sigma$ Dra's age of $4.54 \pm 0.92$ Gyr, the highlighted points in panel (b) show models with $\delta\nu_{02}$ within 2 standard deviations from the star's age and $\Delta\nu$. The $\delta\nu_{02}$ interval spanned by such models is indicated by horizontal lines. This range of predicted $\delta\nu_{02}$ is shown in $\sigma$ Dra's observed collapsed \'echelle diagram in panel (c), which is annotated with mode spherical degrees. The spectrum used to construct this diagram is smoothed using a 3$\,\mu$Hz boxcar filter, while the collapsed \'echelle diagram is background-subtracted and normalized by dividing by the height of the collapsed $l=0$ mode profile. }
    \label{fig:seis_diagrams}
\end{figure}

\subsection{Quadrupole Oscillation Modes}\label{sec:quad}
Our approaches for measuring individual mode frequencies in this study did not identify any significant quadrupole ($l=2$) p-modes in $\sigma$ Dra's TESS oscillation spectrum, nor can they be observed clearly in the \'echelle diagram in Fig. \ref{fig:fig1}b. Such modes greatly improve the age precision obtainable from asteroseismology because their frequencies relative to radial modes provide a measurement of central hydrogen abundance in dwarf stars, which is an indicator of stellar age along the main sequence. Thus, we seek to identify whether any quadrupole modes can be detected in $\sigma$ Dra's oscillation spectrum. 

To determine the expected small frequency spacing ($\delta\nu_{02}$) of the star using a grid of models, we first measured the frequency separation between $l=0$ and $l=1$ frequencies, $\delta\nu_{01} = \frac{1}{2}(\nu_{n, l=0} +\nu_{n+1, l=0}) -\nu_{n, l=1}$, averaged across radial orders ($n$). 
The averaging weight across orders was determined by fitting a Gaussian envelope\footnote{The width of the envelope was determined as $W = \nu_{\mathrm{max}}^k \cdot e^b$, where $k=0.9638, b=-1.7145$ based on parametric fits to the widths of \textit{Kepler} asteroseismic targets \citep{Li_2020}.} centered about a fiducial frequency value of 4250 $\mu$Hz. We measured a value of $\delta\nu_{01}=3.95\pm0.18\,\mu$Hz, which is shown in Fig. \ref{fig:seis_diagrams}a to be consistent with an age of approximately $5\,$Gyr from a grid of stellar models. 

Next, we queried the same grid for all models within 2 standard deviations of $\sigma$ Dra's derived age in Table \ref{table:properties} at the same $\Delta\nu$ and [Fe/H], which yielded models with $11.6\,\mu\text{Hz}\,\leq\delta\nu_{02}\,\leq 14.3\,\mu$Hz as shown in Fig. \ref{fig:seis_diagrams}b. 
This range of $\delta\nu_{02}$ predicts $l\,$=$2$ modes at $(\nu\,\text{mod}\,\Delta\nu)/\Delta\nu \sim0.45$, which is observed to correspond to a slight excess in power as seen in Fig. \ref{fig:seis_diagrams}c. If we assume that the collapsed \'echelle peak height, $\mathcal{H}$, is a proxy for mode visibility, we measure $\mathcal{H}_{l=1}/\mathcal{H}_{l=0}\sim1.15$ and $\mathcal{H}_{l=2}/\mathcal{H}_{l=0}\sim0.2$. These correspond approximately to the expected values as seen in the \textit{Kepler} mission for dwarf stars similar in evolution to $\sigma$ Dra, which are about 1.1 and 0.3 for the relative dipole and quadrupole mode visibility, respectively (c.f. KIC 7970740, \citealt{Lund_2017}).  It is thus plausible that the highlighted regions in Fig. \ref{fig:seis_diagrams}c do indeed correspond to the quadrupole oscillation modes of $\sigma$ Dra.

An important caveat for this identification is that it is based on a smoothed spectrum that is averaged (i.e., collapsed) across radial orders. With an estimated amplitude per radial order of $0.8\pm0.2$ ppm (Section \ref{sec:amplitudes}) and a white noise level of 0.2 ppm as estimated at frequencies above 7000$\,\mu$Hz from the amplitude spectrum, it is likely that individual, low-amplitude quadrupole modes are at the detection limit of the TESS photometry. Meanwhile, EPRV observations can extend beyond this limit to easily detect and measure such modes. As shown for $\epsilon$\,Indi$~$A \citep{Campante_2024}, EPRV observations collected over multiple nights provide
sufficient frequency resolution to resolve $\delta\nu_{02}$, allowing for precise asteroseismic age estimates.







\section{Conclusions and Outlook}

We have detected solar-like oscillations in $\sigma$ Dra using TESS photometry and also with Keck Planet Finder (KPF) in radial velocity (RV). $\sigma$ Dra represents the coolest dwarf star to date for which oscillations are detected in both photometry and radial velocity data. This discovery consolidates the potential of extreme-precision radial velocity (EPRV) instruments for cool-dwarf asteroseismology, while also highlighting the synergy between TESS and ground-based observatories for precisely characterizing the properties of the most promising candidates of future surveys for habitable zone exoplanets from the Habitable Worlds Observatory. 

By modelling oscillation frequencies extracted from TESS photometry, we found an age of $4.54\,$Gyr for $\sigma$ Dra with an age uncertainty of 35\%. The relatively large uncertainty is due to the lack of clear detectability of quadrupole oscillation modes from the TESS data. The oscillations from TESS also show a lack of detectability for high-frequency modes, whereas such modes are clearly visible from the KPF data. However, smoothing and averaging the oscillation spectrum revealed that information regarding $\sigma$ Dra's quadrupole and high-frequency modes may still be recoverable from photometry. Because it is evident that $\sigma$ Dra's oscillation amplitudes are at the limit of detectability in TESS, the KPF data has a clear advantage in unambiguously detecting $\sigma$ Dra's oscillation modes. Thus, additional nights of EPRV observations are expected to yield a more precise age estimate ($\sim$10\,\%) for this star.

With the asteroseismic age estimate having significant potential for improvement, the detailed examination of age-activity relations for $\sigma$ Dra, which is the subject of a follow-up study (Metcalfe et al. in prep), may provide insight into the potential habitability of cool, moderately active dwarfs about the same age as the Sun. In particular, the surface magnetic activity and rotational evolution of these stars place constraints on angular momentum loss through stellar winds \citep{Garraffo_2015}, and consequently how the intensity of the space weather environment may have evolved with time \citep{Metcalfe_2024}. This, in turn, will be vital for assessing the development and retention of atmospheres in the early history of orbiting exoplanets \citep{Airapetian_2020}. Seismically, we find that $\sigma$ Dra's activity levels influence the star's mode amplitudes following a $(L/M)^{1.5}$ scaling relation, similar to the \citet{Campante_2024} result for $\epsilon$ Ind~A. This further supports the recommendation by \citet{Campante_2024} for a calibration of the relation for cool, moderately active dwarfs with $T_{\mathrm{eff}}<5500\,$K. 

With non-contemporaneous photometric and RV measurements, we found that \citet{Kjeldsen_1995} luminosity-velocity amplitude relation predicts values consistent with those observed from $\sigma$ Dra, providing direct evidence that the relation holds for cool dwarfs with $T_{\mathrm{eff}}<5500\,$K. Simultaneous ground- and space- observations of $\sigma$ Dra will allow a detailed study of differences in the properties of solar-like oscillations as seen by photometric and spectroscopic measurements, which have only been permitted for a few stars to date. Such observations are valuable for enhancing our understanding of the dynamics of turbulent convection at the near-surface layers of cool stars (e.g., \citealt{Houdek_1999}), which in turn improves our modelling of mode pulsation properties and predictions of amplitudes in photometry \citep{Chaplin_2005, Zhou_2021}. Because $\sigma$ Dra is fortuitously located in TESS's Northern Continuous Viewing Zone, opportunities for simultaneous observations for $\sigma$ Dra will be abundant, which when combined with a long baseline coverage in the coming years will establish the star as a benchmark target in asteroseismology.

\begin{acknowledgments}
The authors wish to recognize and acknowledge the very significant cultural role and reverence that the summit of Maunakea has within the indigenous Hawaiian community.  We are most fortunate to have the opportunity to conduct observations from this mountain. The data in this study were obtained at the W. M. Keck Observatory, which is operated as a scientific partnership among the California Institute of Technology, the University of California and the National Aeronautics and Space Administration. The Observatory was made possible by the generous financial support of the W. M. Keck Foundation. This research was carried out, in part, at the Jet Propulsion Laboratory and the California Institute of Technology under a contract with the National Aeronautics and Space Administration and funded through the President’s and Director’s Research \& Development Fund Program. Funding for the TESS mission is provided by the NASA Explorer Program. M.H. acknowledges support from NASA grant 80NSSC24K0228. D.H. acknowledges support from the Alfred P. Sloan Foundation, the National Aeronautics and Space Administration (80NSSC21K0652, 80NSSC22K0303, 80NSSC23K0434, 80NSSC23K0435), and the Australian Research Council (FT200100871). T.S.M.\ acknowledges support from NASA grant 80NSSC22K0475. Computational time at the Texas Advanced Computing Center was provided through allocation TG-AST090107. T.R.B. acknowledges support from the Australian Research Council through Laureate Fellowship FL220100117. D.S. is supported by the Australian Research Council (DP190100666). A.C. acknowledges support from the National Aeronautics and Space Administration (80NSSC24K0495). R.A.G. acknowledges the support from the PLATO and GOLF grants of the Centre National D'{\'{E}}tudes Spatiales. T.C.~is supported by Funda\c c\~ao para a Ci\^encia e a Tecnologia (FCT) in the form of a work contract (CEECIND/00476/2018)
\end{acknowledgments}

\facilities{Keck:I (Keck Planet Finder), TESS}

\appendix

\section{Keck Planet Finder Radial Velocity Data} \label{appendix:kpf_rv} 
In Table \ref{table:kpf_data}, we tabulate the measurements of $\sigma$ Dra from the Keck Planet Finder observations as described in Section \ref{section:radial_velocities}.

\begin{table}[ht]
    \centering
    \caption{Radial velocity (RV) time series of $\sigma$ Dra using the Keck Planet Finder. Also reported are the Bisector Inverse Slope (BIS) and Full Width at Half Maximum (FWHM) of the fitted cross-correlation function, and the signal-to-noise ratios of the green (SNR$_{\mathrm{Green}}$) and red (SNR$_{\mathrm{Red}}$) cameras for each timestamp. The full version of this table is available in a machine-readable format in the online journal, with a portion shown here for guidance regarding its form and content. \label{table:kpf_data}}
    \begin{tabular}{ccccccc}
    \toprule
    BJD & RV & RV Error & FWHM & BIS & SNR$_{\mathrm{Green}}$ & SNR$_{\mathrm{Red}}$ \\
    - 2460126 & ($\text{m}\,\text{s}^{-1}$) & ($\text{m}\,\text{s}^{-1}$) & ($\text{m}\,\text{s}^{-1}$) & ($\text{m}\,\text{s}^{-1}$) & & \\
    \midrule
    0.8513017 & -0.3272 & 0.2292 & 6.0847 & -0.0345 & 611.55 & 753.66 \\
    0.8518089 & -0.6665 & 0.2351 & 6.0851 & -0.0355 & 597.33 & 742.41 \\
    0.8523363 & -0.1320 & 0.2494 & 6.0829 & -0.0348 & 563.72 & 703.11 \\
    0.8528543 & -0.1252 & 0.2222 & 6.0824 & -0.0355 & 630.99 & 779.83 \\
    0.8533736 & -0.8242 & 0.2401 & 6.0864 & -0.0359 & 584.74 & 724.46 \\
    0.8538760 & -1.1392 & 0.2171 & 6.0851 & -0.0347 & 644.88 & 790.45 \\
    $\cdots$ & $\cdots$ & $\cdots$ & $\cdots$ & $\cdots$ & $\cdots$ & $\cdots$ \\
    1.1444387 & -1.2487 & 0.4827 & 6.0858 & -0.0337 & 691.98 & 859.92 \\
    \bottomrule
    \end{tabular}
\end{table}

\bibliography{sample631,huber}{}
\bibliographystyle{aasjournal}

\end{document}